\documentclass[
showpacs,
nofootinbib,
amsmath,amssymb]{revtex4}

\usepackage{graphicx}
\usepackage{dcolumn}
\usepackage{bm}
\usepackage{float}
\usepackage[hypertex]{hyperref}
\usepackage{amssymb}
\usepackage{amsmath}

\newcommand{\wtd}[1]{\widetilde #1}

\newcommand{\sss}[1]{\scriptscriptstyle #1}
\newcommand{\Ref}[1]{(\ref{#1})}

\newcommand{\be}{\begin{equation}}
\newcommand{\ee}{\end{equation}}
\newcommand{\leqs}{\leqslant}
\newcommand{\geqs}{\geqslant}
\newcommand{\ba}{\begin{eqnarray}}
\newcommand{\ea}{\end{eqnarray}}

\begin{document}

\title{Complete set of essential parameters of an effective theory}

\author{M.~V.~Iof\/fe}%
\email{m.ioffe@spbu.ru}
\author{V.~V.~Vereshagin}%
\email{vvv@av2467.spb.edu}
\affiliation{Department of High Energy and Elementary Particle Physics, Saint Petersburg State University, Universitetskaya embankment,
7/9, St.-Petersburg, 199034 Russia}



\begin{abstract}
The present paper continues the series [V. V. Vereshagin, True self-energy function and reducibility in effective scalar theories, Phys. Rev. D 89, 125022 (2014); A. Vereshagin and V. Vereshagin, Resultant parameters of effective theory, Phys. Rev. D 69, 025002 (2004); K. Semenov-Tian-Shansky, A. Vereshagin, and V. Vereshagin, S-matrix renormalization in effective theories, Phys. Rev. D 73, 025020 (2006)]
devoted to the systematic study of effective scattering theories. We consider matrix elements of the effective Lagrangian monomials (in the
interaction picture) of arbitrary high dimension $D$ and show that the full set of corresponding coupling constants contains
parameters of both kinds: essential and redundant. Since it would be pointless
to formulate renormalization prescriptions for redundant  parameters, it is
necessary to select the full set of the essential ones. This is done in the
present paper for the case of the single scalar field.

\end{abstract}

\maketitle


\section{Introduction and preliminaries}
\label{Sec_introduction}

In this paper we continue constructing the renormalization scheme
suitable for the single-scalar
{\em effective scattering theory}.
Let us recall that the theory is called effective if the
interaction Lagrangian in the interaction picture contains all the monomials
consistent with the given algebraic (linear) symmetry (see
\cite{WeinEFT}) -- \cite{Weinberg1}).


It is pertinent to note that at present three different concepts are often
confused: 1) effective Lagrangian (should not be mixed with a phenomenological
Lagrangian!); 2) effective field theory (EFT%
\footnote{In what follows we use the following abbreviations:
EFT - for effective field theory; EST - for effective scattering theory,
LSM -- for Linear Sigma Model,  ChPT - for Chiral Perturbation Theory,
RP -- for renormalization prescription}); 3) effective low-energy expansion
(say, ChPT or LSM). All three concepts were first explicitly formulated by
S.Weinberg (see
\cite{WeinEFT}), \cite{Wein2009}, \cite{Weinberg1}
and references therein). Perhaps this circumstance, along with the proximity of the
content of these concepts, is the reason for the aforementioned confusion. Let
us briefly explain the difference between these concepts.

The effective Lagrangian (more correctly, the Lagrangian density) is nothing
but the
\underline{most general}
linear combination of local monomials constructed from the interaction picture
fields (and their derivatives of arbitrary order) that are presented in the
model under consideration. Those terms must be consistent with the Lorentz
invariance and other algebraic symmetry requirements. Clearly, the effective
Lagrangian contains an infinite number of terms, and -- therefore -- an
infinite number of corresponding coupling constants.

The concept of EFT means that it is just Dyson's perturbation scheme based on
the relation
\be
S=T_{\sss W} \!\exp i \!\!\int \!\!  dx L(x)
\label{Dyson}
\ee
(here
$T_{\sss W}$
stands for Wick's
$T$-product
and
$L(x)$ --
for the effective Lagrangian density).
The attractiveness of this concept is explained by the obvious renormalizability
property of EFT: there is a counterterm to absorb the ultraviolet divergency of
any graph. However, this concept also has a shortcoming: it suffers from the
"problem of couplings". One needs to know an infinite number of renormalization
prescriptions (RPs) in order to obtain the well-defined physical predictions
(below we discuss this problem in more detail).

Further, the ChPT is nothing but the calculational scheme adapted for fixing the
coefficients of the amplitude expansion at small values of the mass of the light
boson and its momentum modulo (see
\cite{WeinEFT}, \cite{Gasser-L}). This concept is not as general as EFT just
because the latter does not imply using data to solve the problem of couplings
at every next order of the low energy expansion as it is done in ChPT, which is
"more experimental science"%
\footnote{This expression belongs to J.F.Donoghue
\cite{Donoghue}.}
as compared to EFT (which is a "more theoretical" one). This feature should be kept
in mind when using the term ``effective theory''. EFT claims to describe the
entire kinematics of the amplitudes, while the ChPT is only designed for
describing the low energy region and small mass. These two concepts differ from
each other even by purely technical methods of calculation. When working in any
loop order, the EFT is based on the expression
\Ref{Dyson}.
In contrast, in the cases of ChTP and LSM the expression
\Ref{Dyson}
in each new order uses the Lagrangian (with only few coupling constants)
obtained in the previous stage.

In this paper we study the effective scattering theory (EST) that is just the
effective field theory (EFT) only designed for perturbative calculations of the S-matrix
elements on the mass shell. Green functions may contain divergent contributions:
we are only interested in Dyson's perturbation scheme of calculating the
S matrix%
\footnote{H.Georgi
\cite{Georgi}
called it "on shell effective field theory".}
in the framework of EFT.

Though the effective theory is renormalizable by construction (see
\cite{WeinEFT}, \cite{AVVV2}, \cite{KSAVVV2}
and the references therein) it presents no interest until the "problem of
couplings" is solved. Thus it is necessary to point out an infinite number
of renormalization prescriptions (RPs) that allow one to fix the finite parts
of counterterms. If this is done arbitrarily, the theory loses its predictive
power. Unfortunately, we do not have an infinite number of corresponding
physical principles needed to avoid the problem. Therefore, one must either
indicate new (sufficiently powerful) principles or radically reduce the number
of free parameters in the theory. Anyway, one needs to know how to construct
the whole list of free parameters which
$S$-matrix
depends upon.

For this it looks necessary to understand better the
\underline{general}
features of the EFT construction. The point is that at present the overwhelming
majority of papers in the area of EFT are devoted to the problem of asymptotic
safety in gravity. Since gravity is the very complicated theory, this makes it
too difficult the study of general regularities of EFT. Such a study requires
the systematic investigation of the simplest models -- single-scalar and multiscalar
EFT. Some features were discussed earlier
(see, e.g.,
\cite{Georgi}, \cite{VV2},  \cite{GomisWein}, \cite{KSAVVV2}).
Here we just continue this line: below we consider the
{\em single-scalar effective scattering theory}.

In the recent paper
\cite{VV2}
the part of the problem of couplings has been analyzed by the example of a two-leg
one-loop graph (conventionally called ``self energy''). It was shown that in
order to avoid the problem one needs to revise the concept of one-particle
reducibility/irreducibility (1PR/1PI) and to redefine the notion of ``self
energy''. In this case one only needs to rely upon two conventional physical
conditions (the pole position and the wave function normalization). This
turns out quite sufficient for obtaining the well-defined two-leg function
which can be used as the insertion both in external and internal lines of
Feynman graphs of the effective scattering theory. The possibility of
introducing the alternative definition of reducibility is based on the fact
that the analytical expression for an arbitrary Feynman graph of the effective
scattering theory can be identically rewritten in terms of the
{\em minimal lines}
and
{\em minimal vertices}
by means of the operation called
{\em line reduction}.

It seems pertinent to us to briefly recall the definitions of the above terms.
The effective vertex is called minimal with respect to its line with momentum
$k$
if it does not contain the "killing" factor
$k^2-m^2$
(here
$m$
stands for the particle mass; it is implied that
$m \neq 0$).
The line is called minimal if the adjacent vertex (or both vertices, when the
line in question is internal) is minimal with respect to it. The graph is
called minimal if all of its lines are minimal (these definitions along with
many examples have been suggested in
\cite{AVVV2}
and
\cite{KSAVVV2}).
At last, the parameters that appear in the analytic expression for the minimal
effective vertex are called the
{\em minimal resultant parameters}.

Thus one concludes that the S-matrix elements only depend on the minimal
parameters and thus do not require introducing RPs fixing the nonminimal ones.
In turn, this means that the minimal resultant parameters are the only ones
needed to fix the physical content of the EST.

Another problem waiting for its solution is the following. To calculate graphs
one needs to construct a convenient form of recording of the n-leg effective
Lagrangian monomials in the interaction picture. The point is that the form
suggested in
\cite{VV2}
is not convenient for calculations because it is excessively general: it
contains many identical terms. So, we need to suggest the more suitable form
(without loss of generality) and, hence, to suggest the complete set of
essential coupling constants (see
\cite{WeinAsySafe})
needed to fix the four-leg minimal effective vertices%
\footnote{When writing the term "vertex" we mean the corresponding Lagrangian
monomial (or - simpler - monomial) multiplied by a constant.}.
To put it another way: it is necessary to point out the complete set of basic
four-leg Lagrangian monomials (basis) with certain highest dimension
$D$ (see footnote
\footnote{When using the term "dimension" we mean the "physical dimension"
$D$
calculated in units of
$m$. Keep in mind, however, that, starting from the next Section, we deviate
from this rule and express the physical dimension in units of
$m^2$.}).
The linear span of these basic monomials creates the relevant linear space of
corresponding physical dimension. The coefficients at the individual basic
monomials present the essential coupling constants.

It is the problem of constructing the basic set of monomials (more correctly, of
the corresponding matrix elements in the momentum representation) which we solve
in the given paper. This issue is important in the renormalization theory,
because it would make no sense to formulate the renormalization prescriptions
(RPs) that correspond to redundant parameters.

Below we use the same notations as those in
Ref.~\cite{VV2}:
\be
\partial^{[n]} \stackrel{\rm def}{=}
\partial^{\mu_1}\! \ldots\, \partial^{\mu_n}, \qquad \qquad
\partial_{[n]} \stackrel{\rm def}{=}
\partial_{\mu_1}\! \ldots\, \partial_{\mu_n}\,
\nonumber
\ee
and
\be
\partial^{[a,b,c]} \stackrel{\rm def}{=}
\partial^{\alpha_1,\ldots,\alpha_a;
           \beta_1,\ldots,\beta_b;
           \gamma_1,\ldots,\gamma_c}\, ;
\qquad \qquad \qquad
\partial^{[a,b]}_{[c]} \stackrel{\rm def}{=}
\partial^{\alpha_1,\ldots,\alpha_a;
           \beta_1,\ldots,\beta_b}_{\gamma_1,\ldots,\gamma_c}\, .
\nonumber
\ee
(Latin letters show the number of Lorentz indices which are denoted by Greek
letters.)


\section{Three-leg effective vertex }
\label{Sec_3leg}

In the field theory two Lagrangian vertices (densities) that differ from one
another by surface terms arising from the integration by parts in action are
considered equivalent. In certain cases this allows one to simplify the form of
the vertex record. Let us illustrate this with the simplest example.

Let the three-leg vertex have the form
\be
V = G:\phi \partial^\mu \phi \partial_\mu \phi :\, .
\label{phi_cub}
\ee
Here the sign
$:\ldots:$
stands for the normal product and
$G$
is the dimensional coupling constant (with
$D=1$ in the units of
$m$).
Recall that inside the normal product the field operators commute.

Integrating by parts we obtain the equivalent vertex (as usually, the surface
term is neglected). Since we consider the interaction picture field then
\be
\partial^\mu \partial_\mu \phi = -m^2 \phi\, .
\label{klein}
\ee
This allows one to simplify
\Ref{phi_cub}
as follows:
\be
V = G :\phi \partial^\mu \phi \partial_\mu \phi \!:\,  \simeq \wtd{V} =
-G :\phi\partial^\mu \phi\partial_\mu \phi\!: + G m^2 :\phi^3\!\!: + \ldots, .
\label{equivalence}
\ee
Here we use the symbol
$A \simeq B$
to show the field-theoretic equivalence of the operators
$A$
and
$B$
(the ellipses stand for the neglected term out of integral). From the relation
\Ref{equivalence}
one concludes that the initial vertex
$V$
is equivalent to the triple vertex
$:\phi^3\!:$
$$V =G :\phi \partial^\mu \phi \partial_\mu \phi \!:\,
\simeq \frac{Gm^2}{2} :\phi^3\!: \equiv g_3 :\phi^3\! :
$$
which is more simple than the initial one%
\footnote{Note that the similar trick could not be performed with the vertices
of the  form
\be
V = G^{00n}:\phi \partial^{[n]} \phi \partial_{[n]} \phi:
\nonumber
\ee
where
$n \geqs 2$.
In the latter case one needs to compare the relevant matrix elements -- see
below.}.

Now let us turn to a consideration of the general form of three-leg vertex in
effective theory. The method allowing one to construct the
 $n$-leg
effective vertices was presented in
\cite{VV2}.
It is certainly general. For example, in the framework of this method the
three-leg vertex takes the form
($G^{s_1 s_2 s_3}$
stand for the corresponding coupling constants)
\be
V_3 = \sum_{s_1,s_2,s_3=0}^\infty G^{s_1 s_2 s_3}
    :\left(\partial^{[s_1]}\partial_{[s_2]}\phi\right)
     \left(\partial^{[s_2]}\partial_{[s_3]}\phi\right)
     \left(\partial^{[s_3]}\partial_{[s_1]}\phi\right):\; .
\label{gen eff_3vert}
\ee
The problem is that this form is excessively general because it contains many
identical terms. This makes difficult the interpretation of coupling constants
and complicates calculations.

As suggested in
\cite{VV2}
one can make use of the integration by parts (together with the boundary
condition at infinity) in order to simplify the form
\Ref{gen eff_3vert}.
Taking away the derivatives from the first factor and lowering and raising
indices one can obtain the expression given in
\cite{VV2}.
Further simplification is also possible but it requires comparison of matrix
elements in the space of three-particle states (together with the energy-momentum
conservation low). From this comparison one finds that an arbitrary three-leg
vertex in the interaction picture is equivalent to the simplest one:
$$
D^{s_1 s_2 s_3}
    :\left(\partial^{[s_1]}\partial_{[s_2]}\phi\right)
     \left(\partial^{[s_2]}\partial_{[s_3]}\phi\right)
     \left(\partial^{[s_3]}\partial_{[s_1]}\phi\right)
\simeq g_3:\phi^3:\ .
$$

So, the final form of the effective three-leg vertex reads
\be
V_3 = g_3:\phi^3: \ .
\label{triple_OPvertex}
\ee
Here
$g_3$
is the true resultant coupling constant that has the dimension of mass:
$D(g_3)=D(m)=1$

In what follows we shall often omit the sign of the normal ordering though it
will be implied.


\section{Four-leg effective vertices }
\label{Sec_nleg}

Let us now analyze the more complicated object -- the four-leg effective vertex.
Precisely as above we will use two different principles: (a) Zero boundary
condition at infinity and (b) two vertices are equivalent if the corresponding
matrix elements are the same.

The obviously general form of the four-leg effective vertex reads
\be
V_4 = \sum_{s_1\ldots s_6=0}^{\infty}  g^{s_1\dots s_6}
\left(\partial^{[s_1s_2s_3]}\phi\partial_{[s_1s_4s_5]}\phi
      \partial^{[s_4]}_{[s_2s_6]}\phi \partial^{[s_5s_6]}_{[s_3]}\phi \right)\, .
\label{eff_vert_4leg}
\ee
Let us consider the corresponding Lagrangian monomial with fixed dimensionality
\be
L{(s_1,\ldots ,s_6)} \equiv
\left(\partial^{[s_1s_2s_3]}\phi\right)\left(\partial_{[s_1s_4s_5]}\phi\right)
\left(\partial^{[s_4]}_{[s_2s_6]}\phi\right)\left(\partial^{[s_5s_6]}_{[s_3]}\phi \right)
\label{initial_4leg_monom}
\ee
and make use of the integration by parts in order to simplify it. Clearly,
integrating by parts
$(s_1+s_2+s_3)$
many times one can remove all the derivatives from the first factor. This means
that the expression
\Ref{initial_4leg_monom}
is just a sum of more simple Lagrangian monomials of the form
\be
L^{p,q,r} \simeq L(0,0,0,p,q,r) =
               \phi\partial^{[p]}_{[q]}\phi
               \partial^{[q]}_{[r]}\phi
               \partial^{[r]}_{[p]}\phi\, .
\label{3index_monom}
\ee

So, it is shown that integration by parts allows one to use the three-index
monomials (and, hence, the three-index coupling constants%
\footnote{This form was used in
\cite{VV2},
in that case the problem of identical terms was not important.})
in place of the six-index ones. However, this is not the result that we would
like to get. The point is that the form
\Ref{3index_monom}
still contains identical contributions. Therefore it is necessary to further
simplify the monomials of this kind. Let us do this.

As above, we may reduce the order of derivative operator in the second
factor in
\Ref{3index_monom}:
$p+q \rightarrow p+q-1$.
Integration by parts allows one to get the following relation (the
corresponding calculation is simple but tedious; see Appendix A)
\be
L^{p,q,r} = m^2 L^{p-1,q,r} - L^{p-1,q+1,r} - L^{p-1,q,r+1}\ .
\label{3ind_lowering}
\ee
Thus we can make the first index equal zero and need to consider only the
two-index monomials of the form
\be
L^{0qr} =\phi {\partial}_{[q]} \phi {\partial}^{[q]}_{[r]} \phi {\partial}^{[r]} \phi\ .
\label{2ind_monom}
\ee
These are the monomials which could seem to be the basic ones. Nevertheless,
this is not true. The thing is that the monomials
\Ref{2ind_monom}
are obviously symmetric in their indices:
\be
L^{0qr} = L^{0rq}\  \equiv L^{qr},
\label{2ind_mon_symm}
\ee
therefore only those of them are independent which have
$q \leqslant r$.
If we want (as we do) to work with the monomials that have the
dimensionality
$
D \leqslant N\ ,
$
(in the units of
$m^2$)
then forms
$L^{r, N-r}$
with
$r \leqslant [N/2]$
might be considered independent. Here the symbol
$[x]$
stands for the integer part of
$x$.

At last, considering the monomials with
$q=1$
we see that
$$
L^{1,r} \simeq \frac{m^2}{2}L^{0,r} - \frac{1}{2}L^{0,r+1}.
$$
This means that the monomials of the form
$L^{1,r}$
should be considered dependent and, therefore, excluded from the set of basic
ones.

Does it mean that the most general independent four-field vertex in the effective
Lagrangian of the single-scalar theory can be presented as the linear sum of
terms of the form
\be
V_4 = \sum_{N=0,2,3,...}^\infty  \sum_{q=0,2,3,...}^{E(N/2)}
g_{q,N-q}L^{q,N-q}(\phi)
\label{eff_vert_prelim}
\ee
where
$g_{q,N-q}$
stands for the corresponding independent coupling constants%
\footnote{Only the independent coupling constants present a set of
essential parameters of a theory.}?
The answer is no. The reason is that among different interaction picture
operators
$L^{q,N-q}$
acting on the space of four-particle states
$|k_1,k_2,k_3,k_4\rangle$
(here
$k_i^2=m^2$
and
$k_1+k_2+k_3+k_4=0$
due to the conservation law) there are many dependent ones. We discuss this
point in the next Section.


\section{The space of matrix elements.}
\label{Sec_matr_el_space}

Let us denote
$M^{qr}$
the matrix element of the four-field Lagrangian monomial
$L^{qr}$
between the vacuum and four-particle state
$|k_1,k_2,k_3,k_4\rangle$
under the conditions
$k_1+k_2+k_3+k_4=0$
and
$k_i^2=m^2$.

It is possible to show (see below) that
\begin{align}
M^{22} & \simeq
\frac{1}{2}M^{04}-\frac{4}{3}M^{03}+M^{02}-\frac{1}{18}M^{00}, \label{M22}
\\
M^{23} & \simeq
\frac{1}{10}M^{05}+0M^{04}-\frac{1}{2}M^{03}+\frac{1}{2}M^{02}-\frac{1}{30}M^{00}.
\label{M23}
\end{align}
These relations demonstrate that the corresponding monomials are linearly
dependent and, therefore, the constants
$g_{22}$
and
$g_{23}$
are not independent parameters of a theory.

It might seem that the elements
$M^{0N}$
create a basis in the linear space
$M_N$
of the above-described four-field monomials
$\langle 0|L^{q,r}|k_1,k_2,k_3,k_4\rangle$
with
$q+r=N$.
However, for arbitrary
$N$
this is not true. For example, one can deduce the relation
\be
M^{24}=-\frac{2}{3}M^{33}+\frac{1}{6}M^{06}-\frac{2}{5}M^{05}+\frac{1}{2}M^{04}
-\frac{2}{3}M^{03}+\frac{1}{2}M^{02}-\frac{1}{30}M^{00}
\label{M33}
\ee
which demonstrates that
$M^{24}$
is just a linear combination of another
({\em linearly independent}!)
matrix elements with
$N\leqs6$.
Therefore it must be excluded from the set of basic monomials of the index
$N \geqslant 6$.

We find it pertinent to note that the number
$N$
indicates the physical dimension
$D_N=N$
(in the units of
$m^2$)
of the relevant matrix element
$M^{qr}$
from the space
$M_N$.
This matrix element is a homogeneous polynomial (with the homogeneity degree
$N$)
that depends on three Mandelstam variables
$s,t,u$
$$s=(k_1+k_2)^2,\ \ \ t=(k_1+k_3)^2,\ \ \ u=(k_2+k_3)^2,$$
restricted by the conditions
$s+t+u=4m^2$
and
$k_i^2=m^2$:
\begin{equation}
\begin{split}
M^{qr}(s,t,u)
&
\equiv \langle 0|
\int dx :\phi \partial_q \phi \partial^q_r \phi \partial^r \phi :
|k_1,k_2,k_3,k_4\rangle = {} \\
& {} =
(2\pi)^4\cdot 4 \cdot (-1/2)^{q+r}
\left[
\hat{s}^q
\left(
\hat{t}^r + \hat{u}^r
\right) +
\hat{t}^q
\left(
\hat{t}^r + \hat{u}^r
\right) +
\hat{u}^q
\left(
\hat{t}^r + \hat{s}^r
\right)
\right]\ .
\label{MN_VN}
\end{split}
\end{equation}
Here
$\hat{s} = s-2m^2,\ \ \ \hat{t} = t-2m^2,\ \ \ \hat{u} = u-2m^2$.
In fact, we are more interested in these matrix elements than in the monomials
$L^{qr}$
in coordinate space. The point is that we are studying the effective scattering
theory and thus we need to consider the minimal graphs (see
\cite{AVVV2} and \cite{KSAVVV2})
while the concept of minimality is well defined only in the momentum
representation. Therefore we have to provide the full set of constants that
allow to fix the graph as a function of relevant kinematic variables. It is
more appropriate to do this in momentum space.

In place of three dependent dimensional variables
$(\hat{s},\hat{t},\hat{u})$
it is more convenient to work in terms of three dimensionless ones:
\be
\sigma \equiv \frac{\hat{s}}{2m^2}, \ \ \
\tau \equiv \frac{\hat{t}}{2m^2}, \ \ \
\xi \equiv \frac{\hat{u}}{2m^2}; \ \ \ \
\sigma+\tau+\xi=-1.
\label{shifted_variables}
\ee
This gives:
\begin{equation}
M^{qr}(\sigma,\tau,\xi)
\simeq
m^{2N}(-1)^N
\left[
\sigma^q
\left(
\tau^r + \xi^r
\right) +
\tau^q
\left(
\xi^r + \sigma^r
\right) +
\xi^q
\left(
\sigma^r + \tau^r
\right)
\right]\ .
\label{MNVN_shift}
\end{equation}
Here the inessential numerical factor
$4 \cdot(2\pi)^4$
has been omitted. The dimensional factor
$D_N=N$
turns out separated and the dependence on the kinematical variables is just a
dimensionless three-variable symmetric polynomial with the leading degree
$N=q+r$.

In this Section we will consider not the linear space of Lagrangian monomials
itself but the closely related space
$M_N$
of matrix elements of these monomials with the dimension
$D = N$.
In accordance with that written above we need to reveal the basic elements of
$M_N$;
the relevant coefficients will be precisely the essential parameters that
completely fix the vertex in question. No redundant parameters can appear in
this case.

Let us -- for example -- show how the relation
\Ref{M22}
in this technique can be obtained. For this we need to write down the explicit formulae for
matrix elements of the monomials with, say,
$N \leqslant 8$ (see
\Ref{MNVN_shift}):
\begin{align}
&M^{00}=D_0(+6); \notag
\\
&M^{02}=D_2(-4a_2+2); \notag
\\
&M^{03}=D_3(-6a_3-6a_2+2); \notag
\\
&M^{04}=D_4(+4a_2^2-8a_3-8a_2+2); \notag 
\\
&M^{05}=D_5(+10a_2a_3+10a_2^2-10a_3-10a_2+2); \notag
\\
&M^{06}=D_6(+6a_3^2-4a_2^3+24a_2a_3+18a_2^2-12a_3-12a_2+2); \notag
\\
&M^{07}=D_7(-14a_2^2a_3+14a_3^2-14a_2^3+42a_2a_3+28a_2^2-14a_3-14a_2+2); \notag
\\
&M^{08}=D_8(+4a_2^4-16a_2a_3^2-48a_2^2a_3+24a_3^2-32a_2^3+64a_2a_3+40a_2^2-16a_3-16a_2+2), \notag
\\
&M^{22}=D_4(+2a_2^2+4a_3);  \label{M00-M06}
\\
&M^{23}=D_5(+a_2a_3+a_2^2+2a_3); \notag
\\
&M^{33}=D_6(+6a_3^2+2a_2^3+6a_2a_3); \notag
\\
&M^{24}=D_6(-3a_3^2-2a_2^3-4a_2a_3+a_2^2+2a_3); \notag
\\
&M^{25}=D_7(-3a_2^2a_3-7a_3^2-3a_2^3-6a_2a_3+a_2^2+2a_3); \notag  
\\
&M^{34}=D_7(+a_2^2a_3+5a_3^2+a_2^3+3a_2a_3); \notag
\\
&M^{26}=D_8(+2a_2a_3^2+2a_2^4-9a_3^2-4a_2^3-8a_2a_3 +a_2^2+2a_3); \notag
\\
&M^{35}=D_8(-7a_2a_3^2-2a_2^4-6a_2^2a_3+3a_3^2+a_2^3+3a_2a_3); \notag
\\
&M^{44}=D_8(+8a_2a_3^2+2a_2^4+8a_2^2a_3+4a_3^2),  \notag
\end{align}
where two important combinations
\be
a_2 \equiv \sigma\tau+\tau\xi+\xi\sigma;\ \ \ \ a_3 \equiv \sigma\tau\xi
\label{a2-a3}
\ee
have been introduced. Note that, in accordance with Waring formula (see Appendix B),
the leading terms in all matrix elements are different from zero.

These matrix elements are not the only ones that correspond to all interaction
picture monomials
$L^{qr}$
with
$q+r \leqslant 8$,
$q \leqslant r$.
The clearly dependent ones
$L^{1r}$
(or, better, their matrix elements) have been excluded from our list: as
shown in Sec.%
\Ref{Sec_nleg}
they can be written as linear combinations of the above-presented monomials.

To prove the relation
\Ref{M22}
one needs to write down the general linear dependence/independence condition in
the space
$M_N$
of matrix elements of the vertices with dimension
$D=N=q+r \leqslant 4$:
\be
\alpha_{22}M^{22}+\alpha_{04}M^{04}+
\alpha_{03}M^{03}+\alpha_{02}M^{02}+\alpha_{00}M^{00}=0.
\label{dep_indep_N4}
\ee
Here
$\alpha_{qr}$
stand for the arbitrary numerical factors%
\footnote{The element
$M^{13}$
is absent in accordance with the note above.}.
Now we need to check if this equality can be fulfilled with some factors
$\alpha_{qr}$
that are nonzero. If such a solution exists, we conclude that the elements
$M^{22} \ldots M^{00}$
are linearly dependent. Otherwise they should be considered linearly
independent. In the latter case we can consider them as the basis elements in
the space
$M_4$.

As known (see
\cite{Kurosh}),
an arbitrary
{\em symmetric}
(in variables
$\sigma,\tau,\xi$)
polynomial can be uniquely (identically!) rewritten in the form of an
{\em ordinary}
polynomial depending on three elementary symmetric combinations:
\be
a_1 = (\sigma+\tau+\xi),\ \ \ a_2 = (\sigma\tau+\tau\xi+\xi\sigma),\ \ \
a_3 = \sigma\tau\xi.
\label{sigma123}
\ee
In the case under consideration
$a_1 = -1$,
so, in fact, we deal with the ordinary polynomials
$M^{ij}$
that depend on two combinations
$a_2$
and
$a_3$
only.

The left side of the equation
\Ref{dep_indep_N4}
is nothing but the sum of several ordinary polynomials depending on two
variables
$a_2$ and $a_3$.
To equate this sum to zero one has to equate zero each coefficient of this
sum. This gives one the following homogeneous system of equations%
\footnote{Each equation contains terms corresponding to the fixed dimension
$D \leqslant N$;
the dimensional factors are omitted.}:
\begin{align*}
&2\alpha_{22}+4\alpha_{04}=0; \\
&4\alpha_{22}-8\alpha_{04}-6\alpha_{03}=0; \\
&0\alpha_{22}-8\alpha_{04}-6\alpha_{03}-4\alpha_{02}=0; \\
&0\alpha_{22}+2\alpha_{04}+2\alpha_{03}+2\alpha_{02}+6\alpha_{00}=0.
\end{align*}

This system has a stepped form. The length of the top step equals the number
of terms (in the case in question it is two) with the maximal dimension
$D=4$
in the expressions for
$M^{04}$
and
$M^{22}$
(see
\Ref{M00-M06}).
The thickness of this top step equals to one line.

Let us now move one of the two columns starting with the top line (for example, that
$\sim \alpha_{22}$) from the left to the right side. The matrix of the resulting inhomogeneous system in the left side has triangular form. Thus, its determinant does not vanish, and the system is solvable with regard to $\alpha_{0k},\, k=4,3,2,0.$ Taking $\alpha_{22}=-1,$ one obtains the relation \Ref{M22} with elements $M^{04},M^{03},M^{02},M^{00}$ forming the basis. Alternatively, we could choose four other elements $M^{22},M^{03},M^{02},M^{00}$ as the basic ones. The relation \Ref{M23} can be proved similarly.


The above-considered example, in fact, shows the most important steps, that
should be taken in order to construct a basis in the space
$M_N$
of matrix elements of the dimension
$D \leqslant N$.
It only remains to illustrate the purely technical complications arising when
one passages to the case
$N \geqslant 6$.
For this it is sufficient to analyze the case
$N=6$.
Let us do this.

Consider the linear dependence/independence test:
\be
\alpha_{24}M^{24}+\alpha_{33}M^{33}+\alpha_{06}M^{06}+\alpha_{05}M^{05}+
\alpha_{04}M^{04}+\alpha_{03}M^{03}+\alpha_{02}M^{02}+\alpha_{00}M^{00}=0.
\label{dep_indep}
\ee
With account taken of equations
\Ref{M00-M06}
this equality takes a form of the following system of equations%
\footnote{As above, each equation contains terms corresponding to the fixed dimension
$D \leqslant 6$}:
\begin{align}
(D_6)\ \ [l_1]:\ \ &-2\alpha_{24}+2\alpha_{33}-\, 4\alpha_{06}=0, \notag \\
(D_6)\ \ [l_2]:\ \ &-3\alpha_{24}+6\alpha_{33}+\, 6\alpha_{06}=0, \notag \\
(D_5)\ \ [l_3]:\ \ &-4\alpha_{24}+6\alpha_{33}+24\alpha_{06}+10\alpha_{05}=0, \notag \\
(D_4)\ \ [l_4]:\ \ &+1\alpha_{24}+0\alpha_{33}+18\alpha_{06}+10\alpha_{05}+4\alpha_{04}=0, \label{system} \\
(D_3)\ \ [l_5]:\ \ &+2\alpha_{24}+0\alpha_{33}-12\alpha_{06}-10\alpha_{05}-8\alpha_{04}-6\alpha_{03}=0, \notag \\
(D_2)\ \ [l_6]:\ \ &+0\alpha_{24}+0\alpha_{33}-12\alpha_{06}-10\alpha_{05}-8\alpha_{04}-6\alpha_{03}-4\alpha_{02}=0, \notag \\
(D_0)\ \ [l_7]:\ \ &+0\alpha_{24}+0\alpha_{33}+\,\, 2\alpha_{06} +\,\,\,2 \alpha_{05}+\,2\alpha_{04}+2\alpha_{03}+2\alpha_{02}+6\alpha_{00}=0. \notag
\end{align}
Here the indices
$[l_1]-[l_7]$
stand for the line numbers while
$(D_N)$ --
for the relevant dimensional factors.

The system
\Ref{system}
is stepped and, therefore, solvable. There is no necessity to compute the main
determinant. The only problem is that the thickness of the upper step equals
two lines: the step is too thick. However, this is just an apparent difficulty
that can be easily removed. Indeed, considering the sum of
$l_1$
and
$l_2$
(with corresponding coefficients), one obtains the two new top lines
corresponding to
$D_6$:
\begin{align*}
&-2\alpha_{24}+3\alpha_{33}=0 \\
&-\alpha_{24}+2\alpha_{33}+2\alpha_{06}=0.
\end{align*}
This is precisely what we need.

Now, moving the column with
$\alpha_{24}$
to the right side, we obtain the quite solvable (the main determinant is
nonzero!) system of equations  for
$\alpha_{ij}\,\, (i+j \leqslant 6)$
which shows that
$M^{24}$
is just a linear combination of
$M^{33},M^{06},\ldots,M^{02},M^{00}$.
It can be easily shown that this latter set of matrix elements presents a basis
in the space
$M_6$.

It is clear that the order of the arrangement of columns containing
$\alpha_{24}$, $\alpha_{33}$
and
$\alpha_{06}$
is unimportant. It is possible to bring the system to a stepped form in any
case. For this reason one can consider dependent any one of three elements:
$M^{24}$, $M^{33}$
or
$M^{06}$.
This is just a matter of taste. By our agreement, we always consider
$M^{0r}$
as independent.

The above-considered example can be easily generalized for the case of
arbitrary large
$N$.
Let us formulate the general algorithm how to construct the basis in subspace
of the dimension
$D=N$
under the condition that this has been done already in the subspace with
$D \leqslant (N-1)$:
\begin{enumerate}
\item
Construct the linear combination of
{\em all}
$[N/2]$
different matrix elements
$M^{qr}$
with
$q+r=N,\,\, q \leqslant [N/2]$
and all the
{\em basic}
matrix elements with
$q+r \leqslant N-1$.
\item
Equate this combination to zero. This will give a test of linear
dependence/independence.
\item
Present the obtained test in the form of a system of linear equations for the
coefficients of ordinary two-variable polynomial (with the leading degree
$N$
in the aggregate variable
$\sigma,\tau,\xi$)
depending on two combinations
$a_2(\sigma, \tau, \xi)$
and
$a_3(\sigma, \tau, \xi)$
(see
\Ref{a2-a3}).
\item
The constructed system of equations will take the stepped form. Every step
that corresponds to the dimension factor
$D < N$
will have the thickness equal to one line (the basic elements with
$D < N$
are considered known!). In contrast, the thickness of the top step may happen
to be equal
$t$
lines. The lengths of these top lines will be equal to
$[N/2]$
elements. Since
$[N/2] \geqslant t$
(the total number of all matrix elements
$M^{q,N-q}$
with the index
$N$
cannot be less than the number of linearly independent ones), it is always
possible to rewrite (identically!) the top (thick) step in the completely
stepped form. The length of the finally obtained top step may happen to be
equal, say,
$l+1$.
\item
Move
$l$
columns from the left side of the system to the right side. As a result, one
obtains an ingomogeneous system with nonzero main determinant because the matrix
in the left side is triangular (surely, one has to retain nonzero top step)
\item
The columns corresponding to the coefficients
$\alpha_{qr}$
in the left side will present the basic matrix elements, while those in the
right side -- the linearly dependent ones.
\end{enumerate}

From the above analysis it is clear that the number of independent matrix
elements with index
$N$
is equal to the number of pairs
$[i,j]$
such that
$2i+3j=N$
minus one. A specific choice of a set of these elements is arbitrary.

When the basis in the space of matrix elements
$M^{qr}$
of the interaction picture monomials
$L^{qr}$
with arbitrary fixed maximal index
$N$
is constructed, one can point out the complete set of the corresponding
essential parameters. Those parameters are nothing but the set of numbers
fixing every basic matrix element. As shown above, the basis is just a set
of arbitrary two-variable polynomials that depend on two combinations of
kinematical variables:
$a_2$
and
$a_3$
(see
\Ref{a2-a3}).
This means that we need to point out the numbers fixing those polynomials.
Since an arbitrary polynomial is completely fixed by the values of its
coefficients,
{\em we can consider those coefficients as the only essential parameters of a
theory}.
So, we arrive at the conclusion that the complete set of essential parameters
of the effective single scalar theory is exhausted by the numbers fixing the
coefficients of ordinary polynomials depending on two variables.


\section{Conclusion}
\label{Sec_conclusion}

The effective theory, by definition, deals with the Lagrangian density that
contains all terms consistent with a given linear symmetry. When studying such
theories, we are faced with a problem that was not previously encountered: the
complete list of free parameters that determine the finite
$S$-matrix
is unknown. According to the common belief, this problem does not occur in the
theories with a finite number of terms in the interaction Lagrangian. In this
latter case, all coupling constants, as a rule, are considered as the
independent parameters. One of the important results obtained in this paper shows
that this is not always true. For example, when the dimensions of the
interaction monomials are
$6 \leqslant D < \infty$
(in units of mass squared), one has to take into account only a certain part of
them in order to avoid doubling of some couplings.

One more result has been formulated in the previous Section.
Though it strongly restricts the number of the essential
parameters
of the single-scalar scattering theory only containing the
vertices with {\it finite}
maximal dimension $D_{\rm max}=N \geqslant 6$, the total
number of the essential parameters
of {\em effective} theory (that corresponds to $N \to \infty$)
is actually infinite.
This supports the common belief that the effective theory
itself has no predictive power
(see \cite{WeinEFT} - \cite{Weinberg1}): it requires infinitely
many physical conditions
to formulate an infinite number of renormalization prescriptions
needed to remove the divergences
from the amplitudes of physical processes. In subsequent publication
we will show that this
pessimistic conclusion is wrong. More specific, the unitarity of
the {\it full} $S$-matrix plays here the fundamental role.
A variety of matrix elements contributing to $S_{2\to 2}-$matrix is constructed
from three- and four-leg vertices.
Therefore, the condition of unitarity of the {\it full} $S-$matrix provides
definite relations between parameters
of these vertices. Different terms have
different analytical structure, and
correspondingly, different divergent asymptotic behavior.
Detail calculations show preliminarily that by a suitable
choice of relations between coupling constants and using results
obtained in the present paper, one
can provide unitarity in the case of the single-scalar EFT.
Thus, the principle of unitarity helps
to fix the most part of the essential parameters but several ones.

The third important result is the following. We suggest an explicit algorithm
for constructing the essential bases in that cases when the number of legs of
graphs in question is
$l = 4$
(and, hence, the number of independent kinematical variables is 3). The
algorithm can be generalized for monomials with a greater number of legs.
This note (at least partially) gives an answer to the problems discussed in
\cite{Murayama}.


\section*{Acknowledgements}

V.V.V. is grateful to A.~Drachev, S.~Paston and K.~Semenov-Tian-Shansky for
friendly support. The work of M.V.I. was partially supported by RFBR Grant No. 18-02-00264-a.

\section*{Appendix A: Derivation of Eq.(10)}
\label{Sec_AppendixI}
\mbox

In this Appendix we will prove the relation
\Ref{3ind_lowering}.
Let us write down the left side with the indices
$(p+1,q,r)$.
This monomial must be brought to the form in which the first index is reduced
by one unit. Below there will be many formulas whose essence reduces to
integrating by parts and discarding the boundary terms:
\begin{eqnarray*}
L^{p+1,q,r}&=&
\phi \partial^{[p]\mu}_{[q]} \phi \partial^{[q]}_{[r]} \phi
\partial^{[r]}_{[p]\mu} \phi  =
-\partial^{[p]}_{[q]} \phi
\left[
\partial^{\mu} \phi\partial^{[q]}_{[r]} \phi\partial^{[r]}_{[p]\mu} \phi
+
\phi \partial^{[q]\mu}_{[r]} \phi \partial^{[r]}_{[p]\mu} \phi -
m^2\phi \partial^{[q]}_{[r]} \phi\partial^{[r]}_{[p]}
\right]  =
\\
&=&
\partial^{[r]}_{[p]}\phi \partial_{\mu}
\left(
\partial^{\mu} \phi\partial^{[p]}_{[q]} \phi \partial^{[q]}_{[r]} \phi
+
\phi\partial^{[p]}_{[q]}\phi\partial^{[q]\mu}_{[r]}\phi
\right) + m^2L^{p,q,r}  =
\\
&=&
\partial^{[r]}_{[p]}\phi
\left\{
\partial^{\mu}\phi\partial^{[p]}_{[q]\mu}\phi\partial^{[q]}_{[r]}\phi
+
\partial^{\mu}\phi\partial^{[p]}_{[q]}\phi\partial^{[q]}_{[r]\mu}\phi
\right\}
- m^2L^{p,q,r}+
\\
&+&
\partial^{[r]}_{[p]}\phi
\left\{
\partial_{\mu}\phi\partial^{[p]}_{[q]}\phi\partial^{[q]\mu}_{[r]}\phi
+
\phi\partial_{[q]\mu}^{[p]}\phi\partial^{[q]\mu}_{[r]}\phi
-m^2 \phi\partial^{[p]}_{[q]}\phi\partial^{[q]}_{[r]}\phi
\right\}
+
m^2L^{p,q,r} =
\\
&=&-m^2L^{p,q,r} +
\partial^{\mu}\phi\partial^{[p]}_{[q]\mu}\phi\partial^{[q]}_{[r]}\phi\partial^{[r]}_{[p]}\phi
+\partial^{\mu}\phi\partial^{[p]}_{[q]}\phi\partial^{[q]}_{[r]\mu}\phi\partial^{[r]}_{[p]}\phi
+\partial_{\mu}\phi\partial^{[p]}_{[q]}\phi\partial^{[q]\mu}_{[r]}\phi\partial^{[r]}_{[p]}\phi +
\phi\partial^{[p]}_{[q+1]}\phi\partial^{[q+1]}_{[r]}\phi\partial^{[r]}_{[p]}\phi
\end{eqnarray*}
Thus, due to definition (\ref{3index_monom}) the monomial under consideration
takes the form:
\begin{eqnarray*}
L^{p+1,q,r}&=&-m^2L^{p,q,r}+L^{p,q+1,r}-\phi\partial^{\mu}
\left(
\partial^{[p]}_{[q]\mu}\phi\partial^{[q]}_{[r]}\phi\partial^{[r]}_{[p]}\phi
\right)
-\phi\partial^{\mu}
\left(
\partial^{[p]}_{[q]}\phi \partial^{[q]}_{[r]\mu}\phi \partial^{[r]}_{[p]}\phi
\right)
-\phi\partial_{\mu}
\left(
\partial^{[p]}_{[q]}\phi \partial^{[q]\mu}_{[r]}\phi \partial^{[r]}_{[p]}\phi
\right)
=
\\
&=&-m^2L^{p,q,r}+L^{p,q+1,r}-\phi
\left(
-m^2 \partial^{[p]}_{[q]}\phi \partial^{[q]}_{[r]}\phi \partial^{[r]}_{[p]}\phi
+
\partial^{[p]}_{[q]\mu}\phi \partial^{[q]\mu}_{[r]}\phi \partial^{[r]}_{[p]}\phi
+
\partial^{[p]}_{[q]\mu}\phi \partial^{[q]}_{[r]}\phi \partial^{[r]\mu}_{[p]}\phi
\right)-
\\
&-&\phi
\left(
\partial^{[p]}_{[q]\mu}\phi \partial^{[q]\mu}_{[r]}\phi \partial^{[r]}_{[p]}\phi
-
m^2\partial^{[p]}_{[q]}\phi \partial^{[q]}_{[r]}\phi \partial^{[r]}_{[p]}\phi +
\partial^{[p]}_{[q]}\phi \partial^{[q]}_{[r]\mu}\phi \partial^{[r]\mu}_{[p]}\phi
\right) -
\\
&-&\phi
\left(
\partial^{[p]}_{[q]\mu}\phi \partial^{[q]\mu}_{[r]}\phi \partial^{[r]}_{[p]}\phi
-
m^2\partial^{[p]}_{[q]}\phi \partial^{[q]}_{[r]}\phi \partial^{[r]}_{[p]}\phi
+
\partial^{[p]}_{[q]}\phi \partial^{[q]\mu}_{[r]}\phi \partial^{[r]}_{[p]\mu}\phi
\right) =
\\
&=& -m^2L^{p,q,r}+L^{p,q+1,r}+m^2L^{p,q,r}-L^{p,q+1,r}-L^{p+1,q,r}-L^{p,q+1,r}
+
\\
&+& m^2L^{p,q,r}-L^{p,q,r+1}-L^{p,q+1,r}+m^2L^{p,q,r}-L^{p,q,r+1}.
\end{eqnarray*}
So,
\be
L^{p+1,q,r}=m^2L^{p,q,r}-L^{p,q+1,r}-L^{p,q,r+1}.
\nonumber
\ee
Therefore, step by step the first index of three-leg monomial
$L^{p,q,r}$
can be
lowered down to zero, and we may consider independent only
$L^{0,q,r}.$


\section*{Appendix B: Some properties of symmetric polynomials}
\label{Sec_Appendix2}
\mbox

Here we recall for the reader some facts from the theory of three-variable symmetric
polynomials (see, e.g.,
\cite{Kurosh}:
\begin{itemize}
\item
The three-variable polynomial
$P(x,y,z)$
is called symmetric if
\be
P(x,y,z) = P(y,x,z) = P(z,y,x) = P(x,z,y).
\nonumber
\ee
\item
An arbitrary symmetric polynomial
$P(x,y,z)$
can be presented as just the conventional polynomial depending upon three
symmetric combinations
$$
\sigma_1 = x+y+z;\ \ \ \sigma_2 = xy+yz+zx;\ \ \ \sigma_3 = xyz.
$$
In the theory of three-variable symmetric polynomials this is known as the main
theorem.
\item
Every power sum
$$
S_k \equiv x^k + y^k + z^k
$$
can be calculated step by step with the Newton formula
$$
S_k=\sigma_1 S_{k-1} - \sigma_2 S_{k-2} +  \sigma_3 S_{k-3}
$$
and presented as the polynomial in
$\sigma_1,\sigma_2,\sigma_3$
with the help of so-called Waring formula (see, e.g.
\cite{MacMahon}, \cite{ZengZhou})
:
$$
\frac{1}{k}S_k = \sum \frac{(-1)^{k-i_1-i_2-i-3}(i_1+i_2+i_3-1)!}{i_1i_2i_3}
{\sigma_1}^{i_1}{\sigma_2}^{i_2}{\sigma_3}^{i_3}.
$$
Here the summing runs over all sets of non-negative numbers
$(i_1,i_2,i_3)$
such that
$$
i_1+2i_2+3i_3=k.
$$
\end{itemize}



\begin{thebibliography}{100}


\bibitem{WeinEFT}
 S.~Weinberg,
 Physica {\bf 96A}, 327 (1979).
\bibitem{Wein2009}
 S.~Weinberg,
 {\it Effective Field Theory, Past and Future.}
 arXiv:0908.1964v3 [hep-th], 2009.
\bibitem{WeinAsySafe}
 S.~Weinberg, in
 {\em General Relativity --- An Einstein Centenary Survey},
 ed. by S.~W.~Hawking and W.~Israel
 (Cambridge University Press, Cambridge, England, 1979).
\bibitem{Weinberg1}
 S.~Weinberg,
 {\it The Quantum Theory of Fields}, Vol. 1,
 (Cambridge University Press, Cambridge, 1996).
\bibitem{Gasser-L}
 J.~Gasser and H.~Leutwyler,
 Ann. Phys. (NY) {\bf 158} (1984) 142;
 Nucl. Phys. {\bf B250} (1985) 465.
 \bibitem{Donoghue}
 John F.Donoghue,
 {\it Chiral Symmetry as an Experimental Science},
 CERN preprint CERN-TH.5667/90 (1990).
\bibitem{Georgi}
 H.Georgi,
 Nuclear Physics B361 (1991) 339-350.
\bibitem{AVVV2}
 A.~Vereshagin and V.~Vereshagin,
 Phys. Rev. D {\bf 69}, 025002 (2004).
\bibitem{KSAVVV2}
 K.~Semenov-Tian-Shansky, A.~Vereshagin, and V.~Vereshagin,
 Phys. Rev. D {\bf 73}, 025020 (2006).
\bibitem{VV2}
 Vladimir V.~Vereshagin,
 Phys. Rev. D {\bf 89}, 125022 (2014).
\bibitem{GomisWein}
  J.~Gomis and S.~Weinberg,
 Nucl.Phys. {\bf B469}, 473 (1996).
\bibitem{Kurosh}
 G.~Kurosh
 {\it Higher algebra.}
 (Mir Publishers, Moscow, 1975; translated from Russian).
 \bibitem{Murayama}
 B.Henning et al,
 Commun. Math. Phys. 347 (2016) 363;
 arXiv:1507.07240v1 [hep-th] 2015.
\bibitem{MacMahon}
 P.A.MacMahon.
 {\it Combinatory analysis},
 (ChelseePublishing Co., NY, 1960).
\bibitem{ZengZhou}
 J. Zeng and J. Zhou,
 The Fibonacci Quarterly, 44(2), 117-120 (2006).

\end{thebibliography}
\end{document}